# DSP Based System for Real time Voice Synthesis Applications Development

**Radu Arsinte,Attila Ferencz**-Software ITC S.A.-
109 Gh.Bilascu Street -3400 Cluj-Napoca - Romania
Phone:+40-64-197681,197682  Fax:+40-64-196787
Email:sitc@utcluj.ro
Email:dianaz@utcluj.ro

**Costin Miron**
Technical University Cluj-Napoca - Faculty of Electronics and Telecommunications
26-28 Gh.Baritiu Street ,3400 Cluj-Napoca,Romania

## Abstract

This paper describes an experimental system designed for development of real time voice synthesis applications. The system is composed from a DSP coprocessor card , equipped with an TMS320C25 or TMS320C50 chip, voice acquisition module (ADDA2) ,host computer (IBM-PC compatible), software specific tools.

## 1.Introduction

Theoretical base of digital processing techniques for analogue signals was developed in early '60, for digital simulation of analogue systems and processes[1].

It is possible to implement DSP algorithms an any computer hardware,for exemple a PC, but the rate at which you want to process information determines the optimum hardware platform for your application.There are four categories of hardware widely used for real time DSP implementation.These are:general purpose DSP chips,special purpose DSP chips,bit slice processors and general purpose microprocessors.

In early '80 was launched first monolithic digital signal processing  microprocessor.This sort of device evolved  rapidly , today being capable to execute over 100MIPS .

The rapid advancement of programmable DSP's lets us to satisfy the needs of very demanding applications.The more affordable application for DSP's is voice analysis and synthesis, speed performances allowing execution in real time.

The design of the digital signal processing system must be flexible enough to allow improvements in the state of the art.For rapid prototyping of applications, many DSP producers offer developing systems, going from simple systems as Starter Kits, to more complex like Application Development Boards.The system described is close to the last one , but some facilities make him useful in simple DSP teaching applications.

## 2. System architecture

The system described is builded around (actually inside) an IBM-PC compatible computer , used as host and interface with the user.The block diagram of the system is

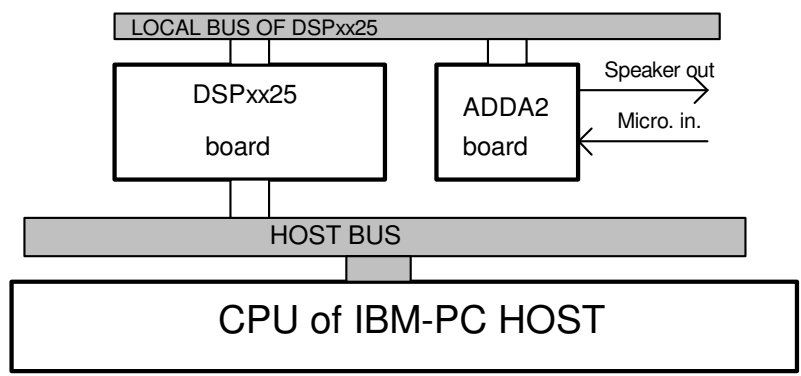

**Figure1. Architecture of the experimental system**

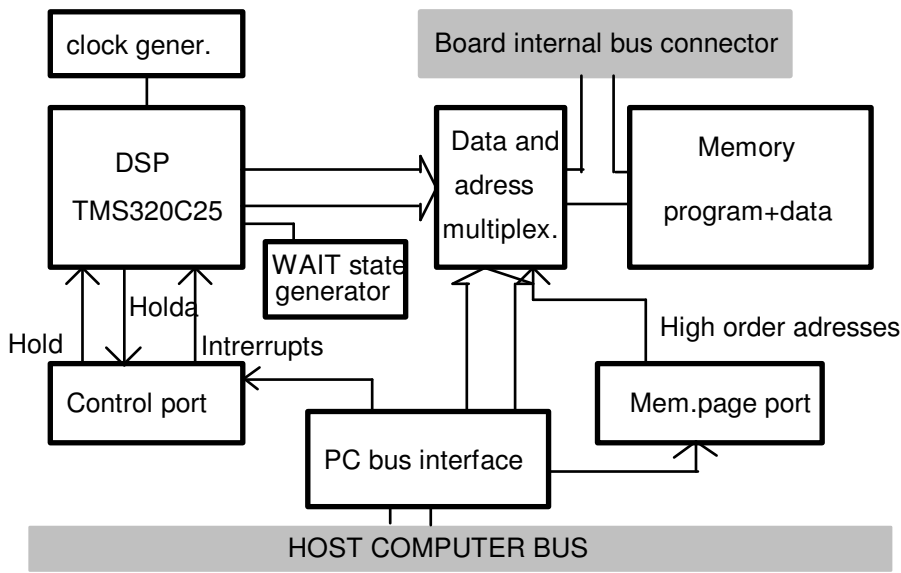

**Figure 2. Block diagram of DSPxx25 board**

described in figure1.

DSP system id designed for applications in various domains of signal processing. For this purpose it has a high degree of modularity, to facilitate adaptation .

The central element is the accelerator board (coprocessor board) called DSPxx25, around him being developed several acquisition modules specialised for various domains: audio frequency, high frequency and video.

The DSPxx25 board is equipped with TMS320C25 fixed point DSP.With a special adapter a TMS320C50 DSP can be used.

The block diagram of the board is presented in figure2.The structure has several particularities.

The memory(data and program) is shared, by multiplexing, between DSP and the host(PC).This feature alows direct access of the host to memory, accelerating the loading of programs and data and extraction of results.Therefore , the transfer speed is close to a memory/memory transfer in PC host, considerably higher than in similar development systems interfaced serial or parallel with the host.

To reduce interference with PC resources, memory from the host side is divided in 'slices' of 8kbytes each.The high order addresses are delivered from an internal special port (page port).

An other port manipulates the control signals of DSP (reset,hold,interrupt request).

For a higher efficiency the board can use one of the interrupt levels of the host .In the application described in [7] this facility is not used.

The performances of the DSP board are the following:
-DSP type                TMS320C25 or TMS320C50
-data format             fixed point
-maximum clock frequency 40MHz
-Wait states             maximum 2
-memory                  64KWords program/data
-extensions              local DSP bus
                         DSP serial
-physical format         AT type board

Acquisition board ADDA2 is coupled on the local bus of the DSP coprocessor board (as seen in fig.1.). The board performs both A/D and D/A conversion and has the following performances:
-Acquisition speed              25KSamples/sec
-Resolution (A/D and D/A)       12 bit
-selectabile amplification between 0.5-3
-polling or interrupt driven acquisition
-3 digital inputs for synchronisation purposes

## 3.Operation

The system has two modes of operation:
1. **Development mode** -designated for the set up of programmes. In this mode the user assembles the application from a library of object files, loads the application in the DSP board and controls the running of program (free-running or step by step) from the integrated debugger.
2. **Autonomous mode**- designated for the final application executing. In this mode the executable program is loaded by the host computer, and runs without any further intervention. Even the reset of the host computer not affect the coprocessor board.

## 4.Software support

Software tools developed especially for this system are:
1. Symbolic assembler, generating directly an executable code (ASMC25)
2. Intelligent debugger for set-up of application(DEPC25)
3. Processing programs library
4. Conversion tools from the Texas Instruments COFF environment

We present shortly few features of them.

**ASMC25** is an symbolic assembler for DSP TMS320C25.This processes source files , written in assembly language an generates object , executable files.The assembler allows the use of directives to control program counter ,memory allocation and the formatting of generated listing.Assemblers functions are:
- syntactical analysis of source file lines
- display of errors detected in assembling process, together with the line number in which the error occurs
- object code generation
- listing file generation (as an option)

**DEPC25** offers the following facilities:
- internal DSP status visualisation
- modification of every internal status element (program counter, accumulator, internal registers, status words, stack)
- interaction with the DSP resources(program memory,data memory):visualisation,altering,constant value filling etc.
- unassembling of desired zones in program memory
- loading of executabile files created with the ASMC25 assembler
- controlled execution of programs (with or without breakpoints,step by step or continuous)

DEPC25 recognises the following commands:
- memory display in hexadecimal format
- unassembled memory content display
- memory move
- memory substitution
- input and output to the desired port
- executable program loading in data or program memory
- global and individual activation/deactivation
- free running execution or execution with breakpoints
- execution resuming after an interrupt or breakpoint
- DSP resetting

**Applications library** contains at this time three types of applications:
- general purpose programs as:FFT,FIR or IIR filters, software interface modules adapted to ADDA2 module
- other useful applications adapted from [2],[3],[4],[5]
- application specific programs as described in [6],[7]

**Conversion tools** are designated to create a bridge between this set of developing programs and the Texas Instruments similar tools (COFF environment). Therefore , we can develop an application in the COFF environment , converting then the COFF file in an object file compatible with DEPC25.

A typical sequence of application development is described in figure 3.

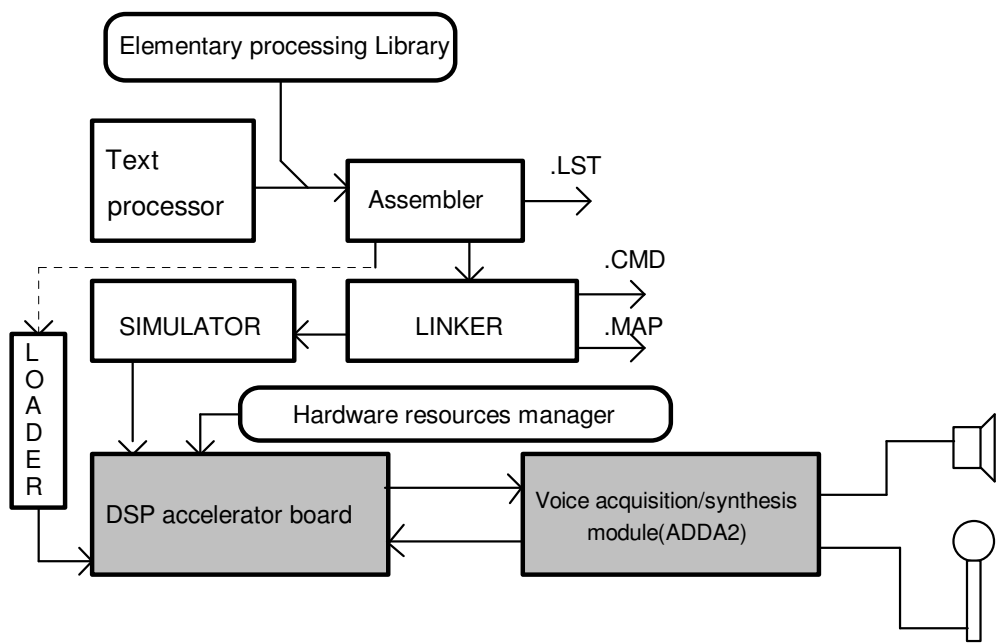

**Figure 3. Sequence of software development in system**

## 5. Conclusion

The development system was used in the real time implementation of speech synthesiser based on the linear prediction method. This application is described in [7].

Using DSP technology allows real-time synthesis of voice, with high quality features. Compared with PC only based systems (without DSP) performances are higher then in a PC486DX4 or Pentium implementation, where is difficult to obtain real-time running with this method(as experiments revealed). On other side in PC throughput limitations occurs, extension bus being considerably slower than the CPU.

After the development phase, done on this system, we can design a dedicated system for consumer applications based on DSP, the resulting system cost being significantly reduced.